# Heterogeneous Dynamic Priority Scheduling in Time critical applications: Mobile Wireless Sensor Networks

Arvind Viswanathan [1], Dr. Garimella Rama Murthy [2]

[1]SASTRA University

arvind20691@gmail.com

[2]IIIT Hyderabad

rammurthy@iiit.ac.in

**Abstract**

In the unlicensed band, the notion of primary user and secondary user (To implement cognitive radio) is not explicit. By dynamic priority assignment we propose to implement cognitive radio in the unlicensed band. In time critical events, the data which is most important, has to be given the time slots. Wireless Sensor nodes in our case are considered to be mobile, and hence make it difficult to prioritize one over another. A node may be out of the reach of the cluster head or base station by the time it is allotted a time slot and hence mobility is a constraint. With the data changing dynamically and factors such as energy and mobility, which are major constraints, assigning priority to the nodes becomes difficult. In this paper, we have discussed about how Wireless Sensor Networks are able to allocate priorities to nodes in the unlicensed band with multiple parameters being posed. We have done simulations on NS-2 and have shown the implementation results.

**Introduction**

Wireless Technologies play an integral role in our day to day lives. Though the channels are reusable, due to increasing number of users, we have to turn to the unlicensed spectrum as there is a limited licensed spectrum. The unlicensed spectrum [1] is one where any user is allowed to access as there is no priority mechanism to allow the more important events to be given higher priorities [2] for communication. This can be disastrous in the long run as the events of higher importance are being starved of the spectrum due to lack of a priority mechanism. We can assume that there are multiple networks competing for the spectrum and hence there is a need to assign a priority for the networks for higher importance during the occurrence of a critical event. Each network itself will have multiple nodes which will also be competing for the channels [3] and will also have to be allotted a certain priority.

To decide how networks are given different priorities and how nodes in each network are allocated priorities we can resort to combined TDMA – FDMA access mechanism.

To reach our goal we need to follow the following steps:

- Select a network
- Select a node

Here we are taking into consideration the following assumptions:

- Mobile base stations, Cluster Heads and sensors (all are mobile)
- Controlled Mobility versus Uncontrolled Mobility

This research paper is organized as follows. In Section I, we shall give a brief background on cognitive radio. In Section II, a discussion on the work previously done is reported. In Section III, we shall discuss about the improvements and the modifications to the existing algorithm. In Section IV and V, we will talk about the new proposed algorithm and the implementations of the algorithm. Finally, we will conclude the paper in Section VI.

## I. Background Study

The concept of cognitive radio is to access the available channel in the wireless spectrum in such a way that more wireless communications are able to run in parallel, at a given point of time, to maximize the spectrum utilization. The spectrum is broadly divided into the licensed and the unlicensed band. The licensed band is further subdivided into primary and secondary users. Primary users are given highest priority and when they are not accessing a channel then it may be allocated to a secondary user. However in the unlicensed band there is no such distinct division between the users and hence any one may use it. Hence, many events of importance such as fire in a building may not be given priority as the priority in the unlicensed band is given randomly. We have used the concept of Cognitive Wireless Sensor Networks [4] in this paper.

## II. Previous Work Done

Modified Distributed Laxity-based Priority Scheduling scheme (MDLPS) [5] is a packet scheduling scheme that improves the average end to end delay and the packet delivery ratio within the deadline when compared with the Distributed Laxity-based Priority Scheduling scheme (DLPS) and the Distributed Priority Scheduling (DPS). The DLPS is a scheduling scheme where the state of the neighboring nodes and the feedback regarding the packet loss from the nodes nearby are taken into consideration.

Previously, priority was given with the help of parameters such as battery power, battery threshold level and mobility in the MDLPS scheme. The Priority Index (PI) was computed with the help of the Packet Delivery Ratio (PDR), Uniform Laxity Budget (ULB), mobility of the node (v), and desired PDR for the flow defined by the parameter M.

$$PI = \frac{PDR}{M} * ULB * \frac{1}{V} * X$$

We have also seen that priority index is high when the priority is low and the priority index is low when the priority is high. Hence, it has an inverse relation with the priority.

$$PI \propto \frac{1}{\text{Priority of the node or network}}$$

The priority of the nodes were computed on the basis of battery power and an associated threshold. An inverse relation was stated because, if the battery power was below a certain threshold, it would not have enough power to send the data. Hence, the nodes which had energy lower than the threshold were given a higher PI. The relationship between Battery Power and the threshold were as follows:

$$BP \propto \frac{1}{PI} \quad \text{if} \quad BP \leq BPth$$

$$BP \propto PI \quad \text{if} \quad BP \geq BPth$$

BPth, is the threshold level of the battery power. The nodes which had full battery power could wait for some time. Hence, they were not given priority first. However, more modifications were done. The number of threshold levels given for battery power were increased. The nodes which were above the hard threshold were divided into multiple levels. This was done so that the nodes having battery voltage greater than the threshold level could avoid undue penalization.

Another important modification that was made, was w.r.t. Packet Delivery Ratio (PDR). When the PDR was below a certain threshold then the packets were discarded by assigning them a higher PI. This is done to allow the nodes with a higher PDR to transmit first so that more data is transmitted.

With respect to mobility the PI had an inverse relationship, i.e. a direct relation with the priority of the node. This was so as the nodes that were moving faster away from the node had to dump the data first so that they do not move away from the range of the decimation node.

$$PI \propto \frac{1}{v}$$

A modification to the calculations in the Universal Laxity Budget (ULB) was also done. The equation was applicable in the case of highly mobile networks. This ensured that the nodes which are about to leave the coverage area of the decimation nodes would dump the data before they left. The new equation was as follows:

$$ULB = \frac{Deadline - Current\ Time}{2^{Remaining\ hops}}$$

### III. Modifications and Improvements

The main objective of the new protocol is to improve in terms of utilizing the unlicensed band efficiently, saving battery power [6], giving priority to the critical events, prioritizing the sensors on the basis of their velocities and data being transmitted. Previously we were trying to improve the end to end delay and the throughput of the system. We did not focus on the need to improve the order of execution of the nodes. The nodes were allowed to execute on the basis of battery power and the velocities. The data values were not taken into consideration which is the most important parameter.

The Mobility [7] was not divided into multiple threshold levels. Mobility (v), has been a factor as the nodes are continuously moving and once they come near the base stations it is necessary that they dump the information. This is so as it is not known when they will be allotted the time slot again and when they will be able to transmit the data. We can segment the mobility into three sections.

- High Mobility ($V_H$)
- Medium Mobility ($V_M$)
- Low Mobility ($V_L$)

Since we are including parameters like data and battery power we have made three different levels. If nodes with data of equal importance and similar battery powers are moving then we will give importance to the node with higher mobility. At the time of the occurrence of a critical event we note the speed of the nodes and assign them to their respective mobility levels. When a priority clash occurs then we will check which level the node is in and assign priorities. This is more efficient than checking it every time a priority clash occurs.

### IV. New Proposed Algorithm

- Different wireless networks and nodes will be represented in the form of priority tuples.

    Priority tuples= (N1, N2)

    N1=Priority index of a network

    N2= Priority index of the nodes in the network of priority N1

    The tuples N1 and N2 can be computed on the basis of the following parameters:

    N1, is decided on the basis of the critical events reported, importance of the data to be communicated, i.e., urgency of the message to be communicated.

    N2, is decided on factors such as, mobility index, i.e., velocity of the node and battery life

Initially, the priority of the network will be selected on the basis of the outreach of the network in the area where the critical event is taking place. In case of a fire in a room, the network which is connected to the room and the rooms near it should be given priority as that network will be able to inform all the nearby nodes about the critical event i.e., the network with higher node density in the area where the critical event has taken place should be given priority. The speed of communication of the network must also be considered. A network with a higher bandwidth should be prioritized.

- Layering and sectoring can be applied to the sensor field to extend the battery life of the nodes. This way the nodes will broadcast only in the required direction thereby saving a lot of power. The nodes with a higher battery power should be given a higher priority index as they will be able to wait for a larger time span.

- There are multiple nodes that want access to the spectrum. The spectrum can be divided with the help of TDMA and FDMA. The intersection points of the TDMA/FDMA will be considered as the positions. We are taking as an assumption that in the unlicensed band, the number of sources are larger than the number of positions. When a particular event occurs the nodes which have been given the highest priority are allocated to the positions. This remains static as long as another critical event does not occur. The nodes can be considered as the sources as they are the ones which get the data and they have to send it to the sinks, i.e. the positions on the TDMA/FDMA frame. The sinks here are the base stations.

- In the case of dynamic scheduling, data that is being transmitted can be used as an important parameter to allot priority. The sensor nodes will be mobile and the base stations will be relatively less mobile. The data will be checked locally by the Cluster Heads (CHs) if it is important. The most important data will be given the lowest priority index, i.e. the highest priority. On a global level all the CHs will check with each other as to which has the most important data to be transmitted and then assign the priorities. Hence, during the occurrence of a critical event such as a fire, the sensor with the most important data in that time slot will be allowed to transmit first to the Base Station (BS). Thus, we can say that the node with the most important data will be given the highest priority.

$$\text{Data Importance} \propto \frac{1}{PI}$$

## V. Implementation

The nodes are given a random way point motion. The cluster heads and the Base Station are also mobile, though their mobility is controlled and is relatively less compared to the sensor nodes. The nodes have random mobility. They are all connected using the UDP agent and a CBR traffic model. We have taken the Two-Ray ground Reflection Model [8] as the nodes are relatively distant from each other. This model is more realistic and is widely used in comparison to the free space model. The terrain area is large and hence this model has been preferred.

The implementation was done on the Network Simulator 2 -2.34 (NS2-2.34) [9]. Some of the parameters taken are given below in the table:

| PARAMETER | VALUE |
| --- | --- |
| Agent | UDP |
| Routing Protocol | AODV |
| Mobility Model | Random Way Point |
| Data Flow | Constant Bit Rate (CBR) |
| Node Placement | Random |
| Terrain Area | 2000 x 2000 |
| Session Duration | 100s |
| Queue Size | 50 |
| Initial Energy Level | 50 |
| Propagation Type | Two-Ray ground Reflection Model |
| Antenna Type | Omni-Antenna |
| Number of mobile nodes | 22 |
| Packet Size | 1000 |

TABLE 1. Parameters taken in the NS2 simulation

The energy model was used to implement a threshold level on the battery. The nodes were assigned different speeds so that the priorities could be allotted on the basis of mobility. The nodes were given an interval of 0.5 seconds between transmissions. The simulation graph and the analytical study of the effect of the new protocol has been given.

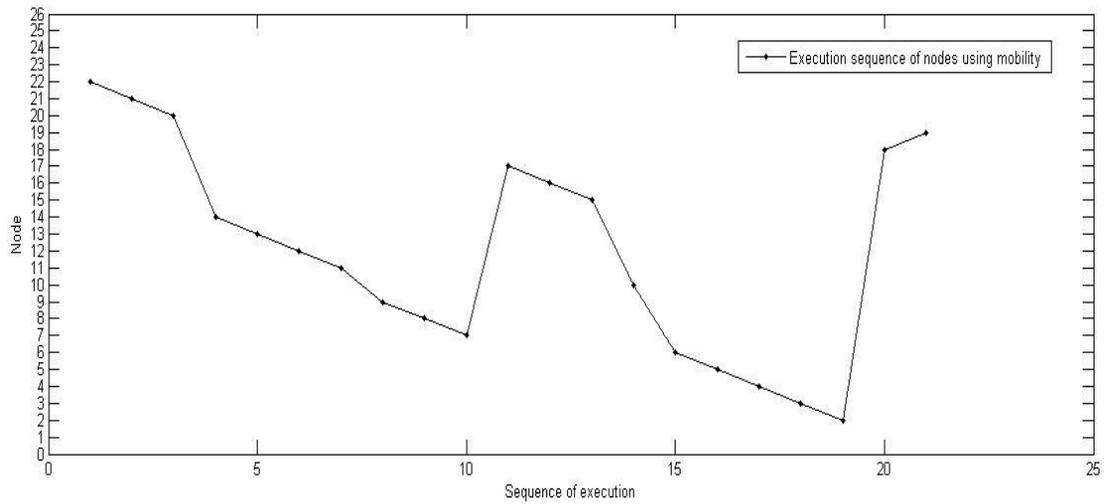

Figure 1. Execution of nodes based on mobility as a priority (MDLPS)

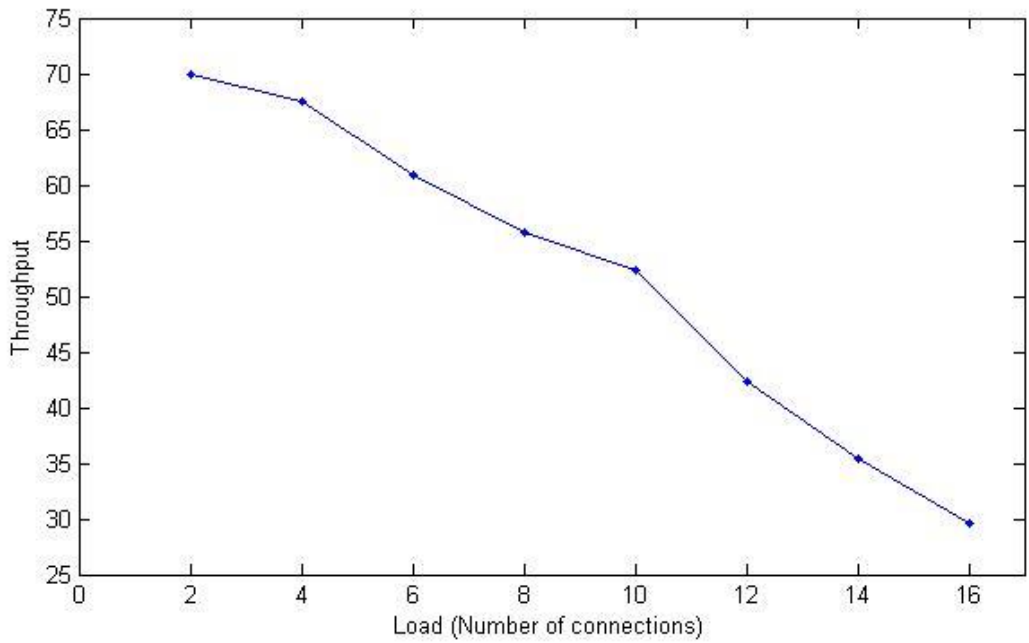

Figure 2. Throughput (kbps) v/s the number of connections

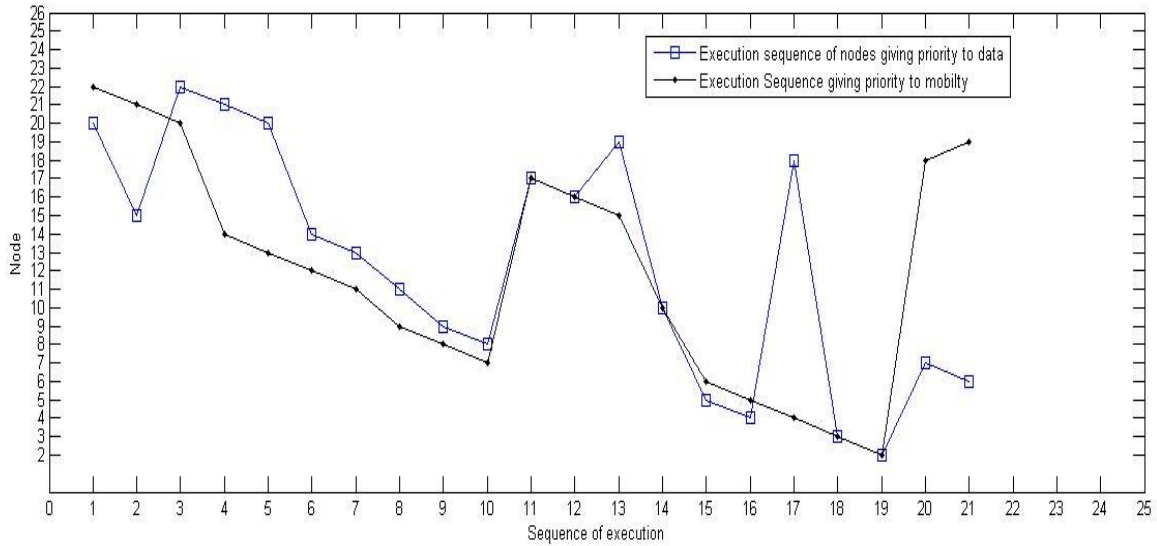

Figure 3. This is a comparison between execution using mobility as a priority (MDLPS) and using data as a priority decider (proposed idea).

Though MDLPS was an improvement over DLPS and DPS, the order of execution was not proper. Hence, the critical events were starved by not being allotted the time slots.

## VI. Conclusion

We have proposed a new algorithm by which we can compute which is the most time critical event. This idea can be implemented in various networks to find out which events are to be executed first in the unlicensed band. The simulation studies have been discussed .We have discussed a few enhancements which could lead to further improve the efficiency of the protocol.

**References**

[1] Ian F. Akyildiz, Won-Yeol Lee, Mehmet C. Vuran, Shantidev Mohanty (2006), NeXt generation/dynamic spectrum access/cognitive radio wireless networks: A survey.

[2] Xue Yang and Nitin H. Vaidya (2002), "Priority Scheduling in Wireless Ad Hoc Networks", Mobihoc 2002.

[3] Thammakit Sriporamanont and Gu Liming (2006), Technical report, IDE0602, Master's thesis in Electrical Engineering, Halmstad University, Wireless Sensor Network Simulator.

[4] Sumit Kumar, Deepti Singhal, and Rama Murthy Garimella (2012), "Cognitive Wireless Sensor Networks" in Intelligent Sensor Networks: The Integration of Sensor Networks, Signal Processing and Machine Learning.

[5] G. RamMurthy, G. Naveen Reddy, A. Ravi Shankar Varma (2011), "Modified Distributed Laxity Based Priority Scheduling Scheme", Wireless Personal Communications, June 2011, Volume 58, Issue 3, pp 627-636.

[6] Dhiraj Nitnaware and Ajay Verma, Energy Constraint Node Cache Based Routing Protocol for AdHoc Network, International Journal of Wireless & Mobile Networks (IJWMN), Vol.2, No.1, February 2010.

[7] Nicolas Eude, Bertrand Ducourthial and Mohamed Shawky (2005), "Enhancing ns-2 simulator for high mobility ad hoc networks in Car-to-Car communication context", The 7[th] IFIP International Conference on Mobile and Wireless Communications Networks, MWCN 2005.

[8] Wu Xiuchao (2004), "Simulate 802.11b Channel within NS2", http://cir.nus.edu.sg/reactivetcp/report/80211ChannelinNS2_new.pdf

[9] The Network Simulator – NS-2. http://www.isi.edu/nsnam/ns.